\documentclass[conference]{IEEEtran}
\IEEEoverridecommandlockouts
\usepackage{cite}
\usepackage{amsmath,amssymb,amsfonts}
\usepackage{algorithmic}
\usepackage{graphicx}
\usepackage{textcomp}
\usepackage{xcolor}

\def\BibTeX{{\rm B\kern-.05em{\sc i\kern-.025em b}\kern-.08em
    T\kern-.1667em\lower.7ex\hbox{E}\kern-.125emX}}
\begin{document}

\title{Data Breaches in Healthcare Security Systems\\
%{\footnotesize \textsuperscript{*}Note: Sub-titles are not captured in Xplore and
%should not be used}
%\thanks{Identify applicable funding agency here. If none, delete this.}
}

%\author{Anonymous for reviewing process
%}

\author{\IEEEauthorblockN{Jahnavi Reddy$^*$, Nelly Elsayed$^*$, Zag ElSayed, Murat Ozer}
	\IEEEauthorblockA{\textit{School of Information Technology} \\
		\textit{University of Cincinnati}\\
		Cincinnati, Ohio, United States\\
		reddyji@mail.uc.edu,nelly.elsayed@uc.edu,elsayezs@ucmail.uc.edu,ozermm@ucmail.uc.edu}
	{\footnotesize \textsuperscript{*}Note: authors have the same amount of contribution.}
	}
	
%\author{\IEEEauthorblockN{1\textsuperscript{st} Given Name Surname}
%\IEEEauthorblockA{\textit{dept. name of organization (of Aff.)} \\
%\textit{name of organization (of Aff.)}\\
%City, Country \\
%email address or ORCID}
%\and
%\IEEEauthorblockN{2\textsuperscript{nd} Given Name Surname}
%\IEEEauthorblockA{\textit{dept. name of organization (of Aff.)} \\
%\textit{name of organization (of Aff.)}\\
%City, Country \\
%email address or ORCID}
%\and
%\IEEEauthorblockN{3\textsuperscript{rd} Given Name Surname}
%\IEEEauthorblockA{\textit{dept. name of organization (of Aff.)} \\
%\textit{name of organization (of Aff.)}\\
%City, Country \\
%email address or ORCID}
%\and
%\IEEEauthorblockN{4\textsuperscript{th} Given Name Surname}
%\IEEEauthorblockA{\textit{dept. name of organization (of Aff.)} \\
%\textit{name of organization (of Aff.)}\\
%City, Country \\
%email address or ORCID}
%\and
%\IEEEauthorblockN{5\textsuperscript{th} Given Name Surname}
%\IEEEauthorblockA{\textit{dept. name of organization (of Aff.)} \\
%\textit{name of organization (of Aff.)}\\
%City, Country \\
%email address or ORCID}
%\and
%\IEEEauthorblockN{6\textsuperscript{th} Given Name Surname}
%\IEEEauthorblockA{\textit{dept. name of organization (of Aff.)} \\
%\textit{name of organization (of Aff.)}\\
%City, Country \\
%email address or ORCID}
%}

\maketitle

\begin{abstract}
Providing security to Health Information is considered the topmost priority compared to any other field. After digitalizing patients records in the medical field, the healthcare/medical field has become a victim of several internal and external cyberattacks. Data breaches in the healthcare industry have been increasing rapidly. Despite having security standards such as HIPAA (Health Insurance Portability and Accountability Act), data breaches still happen on a daily basis. All various types of data breaches have a similar harmful impact on healthcare data, especially on patients' privacy. This paper aims to understand the aspects that led to healthcare data breaches via ransomware incidence and their impact on the patients and healthcare providers. In addition, the paper reviews the current possible solutions to improve the healthcare security system by analyzing the efficiency of these solutions. We studied the most significant healthcare data breaches via ransomware attacks that occurred in the U.S. from 2015 to 2020. We analyzed the obtained data from different academic and business sectors resources that target the reasons for the healthcare data breaches. 
\end{abstract}

\section{Introduction}
Data privacy and security are the topmost priorities, especially in the healthcare field. Since the patients' records are replaced with electronic health records (EHR) from the paper-based system, data breaches events have increased rapidly. There are different types of data breaches occurring, which can be classified as internal and external breaches~\cite{eling2017data}. Lack or poor infrastructure, software vulnerabilities, and unauthorized access to the database can be categorized as internal breaches. Hacking, Malware/Ransomware attacks, and theft are categorized as external breaches. According to a data security study survey~\cite{seh2020healthcare}, the data breaches are as follows: caused by human error $(33.5\%)$, misuse of data $(29.5\%)$, theft $(16.3\%)$, Hacking $(14.8\%)$, Malware $(10.8\%)$. These data breaches affect in many ways. 
\begin{figure}[h]
	\centering
	\includegraphics[width=8.5cm, height = 4.5cm]{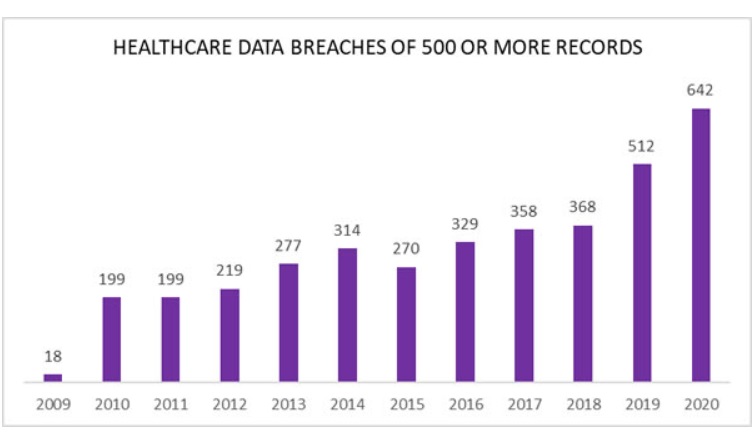}
	\caption{Healthcare Data Breaches of 500 or more records~\cite{7455821}.\\
		Fig.~\ref{fig:s1} Alt Text: A bar diagram of the healthcare data breaches from 2009 to 2020. The diagram shows a significant increase annually in the healthcare data breach cases.}
	\label{fig:s1}
\end{figure} 
Figure~\ref{fig:s1} shows the data breaches between the years 2009 and 2020. Nearly $3,705$ data breaches have occurred with more than 500 records. Observing the pattern, we can clearly state that the number of breaches has been increasing over time except in 2015.

\begin{figure}[h]
	\centering
	\includegraphics[width=8.5cm, height = 5cm]{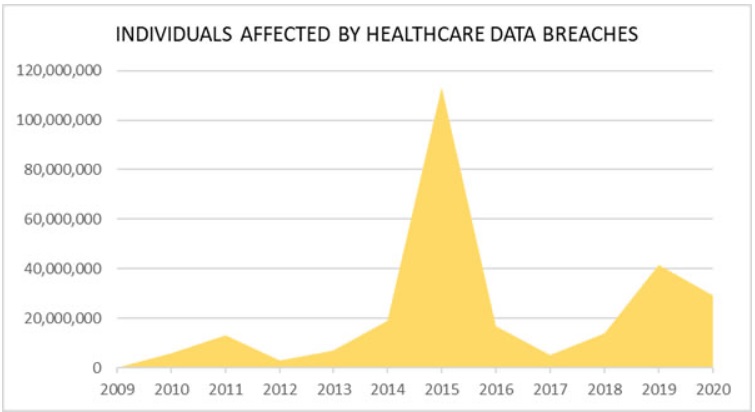}
	\caption{Individuals affected by healthcare data breaches~\cite{Alder2021}\\
		Fig.~\ref{fig:s2} Alt Text: A diagram shows the number of individuals that are affected by healthcare data breaches from 2009 to 2020. The diagram shows that the major healthcare data breaches occured in 2015 and 2020.}
	\label{fig:s2}
\end{figure} 
Figure~\ref{fig:s2} shows the data from 2015, which is the worst year for data breaches, as nearly 113 million data have been exposed. Also, 2015 is considered the most damaging year possible because there were three colossal data breach attacks in the healthcare industry: Excellus, Anthem Inc, and Premera Blue Cross.
These breaches have resulted in the disclosure of $268,189,693$ health records.
During a data breach, all the patients' information has been stolen by hackers~\cite{7455821}. All this information is enough for any individual to take a loan on a patient's behalf. Further, these hackers sell this information on the Dark Web for individuals to perform illegal activities such as buying drugs, financial schemes, or even claiming false insurance~\cite{adewopo2022deep}. All these data breaches are happening despite having standards such as HIPAA~\cite{4753677}, and there are also organizations such as HITRUST. HITRUST is an organization that assists other companies to help them follow the HIPAA standard~\cite{Seh2020}.

\section{Topmost Ransomware Attacks in the Healthcare}

\subsection{Trickbot}
Trickbot is created by Wizard Spider cybercriminal group. Trickbot has several names, such as UNC1878 or Team 9~\cite{celik2019behavioral}. These groups of hackers deliberately target the US hospitals, health, and public sectors. Trickbot created a tool called anchor-DNS which acts as a botnet. It provides backdoor access to infected computers and later uses this backdoor as a gateway to infect other computers using the computer already infected with the malware.
TrickBot uses four main methods to inject malware into the systems: SpearPhishing, Secondary Payload, Network Vulnerabilities, and Malvertising.

\subsection{RYUK Ransomware}
RYUK ransomware is mainly used for financial gain. It is initiated by a cybercriminal group that has injected ransomware in approximately $400$ healthcare facilities in the US~\cite{hassan2019ransomware}. RYUK ranks third in all the ransomware attacks that have happened in 2020. The FBI has claimed that the victims paid approximately $61$ million dollars of ransom to recover the stolen data. 

The RYUK Ransomware works as follows: In the first stage, it attempts to delete all the files and also the backup data. Later it will shadow copy these files and attempts to end all the security services linked with the device. In the second stage, it disables the built-in Windows Automatic Startup Repair. In the third stage, the boot status policy is changed to ignore all failures. In the last stage, a note is left on the screen telling the victim to pay a ransom amount of money to recover all the data and warns the victim not to turn off their device. 
The files named RyukReadMe.html or RyukReadMe.txt, which are created on the victim's desktop, contain the email address to which the ransom payment has to be sent.

A similar attack has also happened in the past, ``The WannaCry Virus," in 2017. Approximately $40\%$ of the health organizations were affected~\cite{ehrenfeld2017wannacry}.
%This ransomware can be infected to hospitals or other medical organizations by various means. The hackers do much groundwork before they perform any hack on an organization. 

\section{Reconnaissance of Hospital Network Occurrence}
\subsection{Footprinting}

Footprinting is the first step of the hacking process. It is also called the analysis step~\cite{levy2021introducing}. Hackers learn all about the targeted hospital, including the working personnel and the computing devices used, including the WiFi, the operating system (OS), and the location of all the computing devices. Using all these details, hackers determine the vulnerabilities of those devices.
\begin{enumerate}
	\item \textbf{Scanning devices:} Hackers buy the same devices from the market. These security devices come with the instruction manual where all the credentials like account name and password will be given away. Before hacking the actual devices, hackers try password cracking on these sample devices. Then, they perform several hacking attacks.
	\item \textbf{Account Harvesting Attack:} It uses a computer program to gather information regarding the staff from the internet from various sources. It collects all the information like phone numbers, email addresses, and other important stuff. Later the hacker uses data mining techniques to analyze this data.
	\item \textbf{Social-engineering attacks:} In this attack, the hacker impersonates one of the suppliers and tries to get information~\cite{krombholz2015advanced}. They send requests to the staff members on various social networking platforms such as LinkedIn and Facebook to get their personal information, such as birthdays. They closely monitor whom the staff members are meeting, their promotion details. 
	\item \textbf{Behaviour-Monitoring Hacks:} Hackers monitor the behavior of the staff and their activities, attempting to implement shoulder surfing for the passwords, personal identification numbers (PINs), and other security codes~\cite{spagnoli2018behaviour}. Sometimes if the hacker cannot physically go to the hospitals, the hacker may install Trojan software like PlaceRaider, which can print 3D models of the targeted hospitals. PlaceRaider can be used to take photos of computer screens, protected health information (PHI), and other financial documents.
\end{enumerate}

\subsection{Scanning}
During Scanning, hackers mostly use the ``Shodan database"~\cite{shohan} to locate the targeted hospital. If the targeted hospital is found in the Shodan database, the hackers can find the OS used by the hospital and the IP address. Other hacking tools are used like Nmap, SNORT, McAfee, Google Hacking Diggity Project, and Nessus~\cite{barnett2008towards}.
%\begin{enumerate}
	%	\item \textbf{Probing Attack:} Once the hacker knows which OS the organization is using, they find out the vulnerabilities in that software because of which the hospital's network is exploited~\cite{handschuh1999probing}.
	%	\item\textbf{Enumeration:} The hacker tries to access the network's config.bog file, which has all the information regarding the system's configuration such as usernames and passwords. It has access to OS accounts which are used for controlling the systems.
	%\end{enumerate}

%\item Spyware Attack- which is secretly installed in the system's network which is used to send information to the hackers without the consent of the hospitals. Spyware can collect all sorts of information from the computers such as personal data, user logins, patient's data, username, passwords, internet habits. Some spyware can also change the systems software settings.
%

\subsection{Network Mapping}
This is the last step, the hacker already has a clear understanding of all the devices that the hospital is using, and the devices are misconfigured by the hackers so that they can closely monitor all the activities~\cite{papadopoulos2014network}. These devices become vulnerable and can be easily hacked using the ``pivot points".

\section{Effectiveness of the Current Approaches to Detect Malware}
Most healthcare organizations depend on the discovery and detection of malware. However, detection-based approaches cannot detect all kinds of malware, like the zero-day malware or the malware hidden inside an encrypted file. Delay in discovering such malware leads to massive data breaches and cyberattacks like the WannaCry attack~\cite{mohurle2017brief}.
In recent studies, the research reveals that hackers use a separate encryption method to bypass the detection. So, once the detection fails, they have access to the web browsers of that particular organization. These web browsers are easy to access. That is why the hackers use these browsers like Edge, Firefox, Chrome as a backdoor.
Even though healthcare organizations are using Multi-Factor Authentication (MFA)~\cite{ometov2018multi} to ensure that only authorized people can have access, the data can still be hacked because the malware has already infected the organization’s system through the web browser. MFA cannot protect healthcare organizations from a data breach when the breaches are caused because of insider errors~\cite{das2020mfa,ometov2018multi}.
Most of these threats are out of sight because they are hidden in the storage services such as Google Drive, Cloud, and Dropbox. Due to that, SSL-based threats increased by 260\% in 2020, and almost two billion threats were declared against the healthcare industry.

There are two major methodologies to detect malware or any intrusion:
\begin{itemize}
	\item \textbf{Intrusion Detection System:} The IDS provides security at the office network perimeters. A firewall provides integrated, inline security services and lock-tight security and control for each protocol traversing the office router~\cite{wack2002guidelines}. IDS is recommended to run on all office perimeter interfaces, but tuning may be required to prevent oversubscribing IDS monitoring capabilities.
	
	\item \textbf{Intrusion Prevention System:} The IPS acts as an inline intrusion detection sensor, watching packets and sessions flow through the router, then scanning each packet to match any IPS signatures. When it detects suspicious activity, it responds before network security can be compromised and then logs the event. 
\end{itemize}

To prevent the hospitals from attacks and to prevent unauthorized access to the hospital data, all organizations should follow a set of security standards for security.

\section{Security Standards in Healthcare}
\subsection{HIPAA}
The Healthcare Insurance Portability and Accountability Act came into law in 1996 to make sure that the employees received the health insurance coverage~\cite{eddy2000effect}. In the following years, several changes were made in the HIPAA, which led to the HIPAA Privacy and Security rules. HIPAA  has four rules~\cite{9167635}: HIPAA Privacy Rule, HIPAA Security Rule, HIPAA Breach Notification Rule, and HIPAA Enforcement Rule.

\subsection{ISO 27001/ISO 27799}
ISO 27001 and ISO 27799 are the two international security standards used for the protection of sensitive information in the healthcare sector~\cite{brenner2007iso}. ISO 27001 is a security standard that establishes the information security management systems, whereas ISO 27799 provides the security controls~\cite{ngqondi2009iso}. It includes the list of all the potential threats which need to be addressed by the security management system~\cite{Lexigram}.

\subsection{HITECH}
The Health Information Technology for Economic and Clinical Health Act makes it mandatory to provide the security of protected health information (PHI) as the topmost priority~\cite{Lexigram}. HITRUST CSF is an organization that maps all the information from different standards like ISO 27001, NIST, HIPAA, HITECH and helps the other healthcare organizations to achieve their security compliance~\cite{Lexigram,ibrahim2018security}.

\subsection{NIST CSF}
Healthcare organizations are at constant threat of different cyberattacks. To address all such attacks, NIST has developed a Cybersecurity Framework (CSF). The NIST CSF is a structured framework built to perform risk analysis and detect emerging threats to the organization.

%\subsection{HITRUST}
%\subsection{DoD 8500}
%\subsection{NIST RMF}

%\section{Problem statement}
%Since the patient's records are replaced with EHR from the paper-based system, fear of unwanted data exposure has increased. The government has taken various measures to prevent data breaches; organizations like HITRUST make sure that the healthcare companies are following the security standards such as HIPAA and HITECH. Since they have also increased the funding for the security of the data, many breaches still happen every day despite all these measures. The number of data breaches that are occurring has increased rapidly, especially in recent times due to the global pandemic, COVID-19, which causes a significant problem as much sensitive information such as first name, last name, Social Security Number (SNN), phone numbers, financial, and insurance information are being leaked through these breaches and this information is sold for approximately $\$175$ per each record costing the companies about $3.86$ million dollars per each data breach.
%To reduce future breaches, the front-line workers should be able to track the devices that have sensitive data. Also, instead of storing the sensitive information in all the devices and giving everyone access, restricting the number of systems that have the sensitive information could result in less potential harm.

\section{An Examination of Threats and Enforcement in the Healthcare Sector}

\begin{figure}[h]
	\centering
	\includegraphics[width=8.5cm, height = 4.5cm]{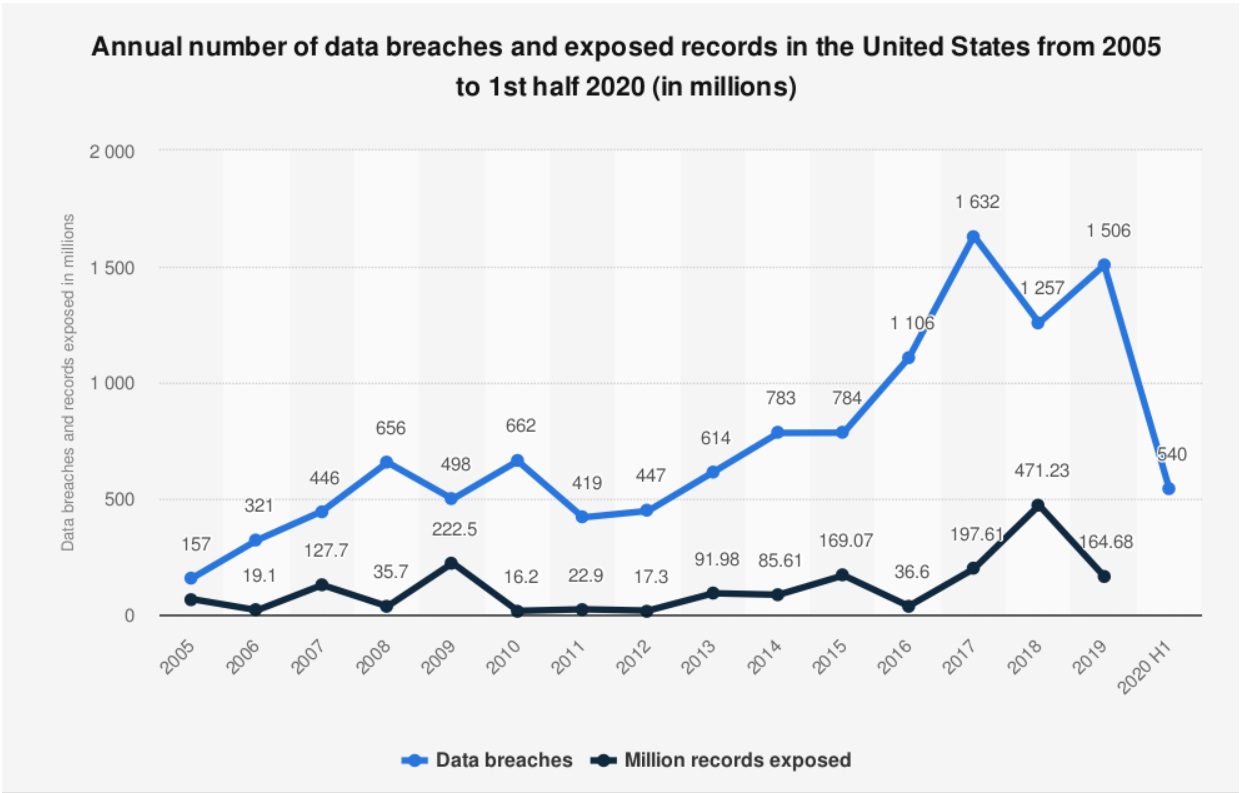}
	\caption{Number of data breaches and exposed records.\\
		Fig.~\ref{fig:s3} Alt Text: A diagram that show the annual number of breached and exposed records in the United States from 2009 to 2020. The diagram indicates that the number of data breaches is higher that the exposed records through the entire time period.}
	\label{fig:s3}
\end{figure}

Most industry members consider HIPAA inefficient and insufficient because of the lack of in-depth technical requirements and specifications. It has compromised most of the patient's privacy during most of the data breaches. Figure~\ref{fig:s3} shows the number of healthcare data breaches of more than 500 records from 2009 to 2020.
After examining the history of ransomware, the researchers have concluded that every attack that has happened is different, and the malware is evolving with time and is becoming more undetectable. Some providers warn that no matter how small the healthcare organization is, it is still prone to a cyber-security attack, and no organization is immune to these attacks. This is because these organizations follow only the essential security controls that are very easy to pass through. The last time HIPAA had any significant changes was in the year 2013. Even though many attacks had occurred in significant years, there was no attempt to change these regulations or update them, putting the patients' sensitive data at risk. HIPAA security rule allows the entities to evaluate their infrastructure and requirements and implement their specifications as required. If the organizations do not want a particular rule not to be implemented, they can justify why they do not want to implement that particular security rule. So, there is flexibility to customize their cybersecurity implementations, resulting in data breaches by putting the health organizations at risk.
HIPAA does not specify the levels of security required for the organizations. The only two rules that are mandatory in HIPAA are the emergency access procedures and standard access controls. The other security controls are addressable and are optional, like encryption. So most companies/organizations do not encrypt the EHR and have only a basic level of security. Therefore, proficient, skilled hackers can easily hack through the EHR, which is not encrypted. If higher security and encryption techniques are implemented, the organizations would be less likely to be attacked.

\section{Reasons of Frequent Cyber Attacks on Heathcare Systems}
The health systems are made up of applications and products that third parties develop. Hence, the products might come from different organizations. If those third-party organizations are not secure enough, there is a possibility that this might affect the healthcare organization using their products. Another barrier discussed was one often seen in Healthcare Preparedness, limited funds, and limited time. There are different IT solutions for each cyberattack, and each solution might come with a different cost. Therefore, it becomes difficult for small-scale organizations to pay such a massive amount of money for security. Also, small healthcare organizations do not have an emergency security advisor in case of an attack or a data breach. It is always a secondary job for some other duty~\cite{Rock2020}.
Due to that, many organizations have been affected in recent times where many sensitive data from the hospitals has been compromised. The topmost organizations that were affected by the security attacks are listed below.

\section{Significant Security Breaches in Healthcare}
\subsubsection{UVM Healthcare Breach}
The University of Vermont Health Network was disrupted by a cyberattack which was done by the Russian Group of Hackers "Ryuk"~\cite{byrne2021cybersecurity}. This attack involved Ransomware~\cite{hassan2019ransomware,kusuma2021network}. The patients' data was held hostage; there was no phone call or any letter on how to retrieve all this data. However, an anonymous file was deposited in the system on how to contact the attackers. This attack occurred on October 28 and lasted for almost forty-two days. The hospital IT staff noticed the abnormal activity and shut down its online operations. They hacked down hospital phone and email networks, eliminated access to patient records and appointments, and halted access for the hospital to pay employees overtime. The revenue lost during this attack could exceed \$63 million. The attackers encrypted files and data behind "virtually all" of the UVM Medical Center's servers — about 1,300 in total — and deposited malware onto more than 5,000 computers and laptops. After three days after the attack had happened, the IT staff were able to access only the read-only files like medical records, meaning that nurses and doctors could view patients' medical histories, prescriptions, and past appointments.
\subsubsection{Trinity Healthcare Breach}
The Trinity Healthcare Breach has affected nearly $3,320,726$ individuals. In July $\mathrm{16^{th}}$ 2020, Blackbaud informed Trinity Health that a cyberattack involving ransomware occurred, which has impacted donor database backup files maintained by Blackbaud, including Trinity Health’s donor database. When Blackbaud was notified regarding the attack on Trinity Health, they conducted a self-investigation to check if any sensitive information was compromised. After the investigation, some information like donor’s details like name, address, phone number, email address, most recent donation date, age, date of birth, and dates of service might have been compromised. Most of the files, such as the social security number (SSN) and other confidential details, were encrypted; hence the cybercriminals could not access this data. This database which was leaked, was from the year 2000 to 2020. This attack was reported to have occurred between April $\mathrm{18^{th}}$, 2020 - May $\mathrm{16^{th}}$ 2020. The Blackbaud took immediate steps to stop the cyberattack by locking out the cybercriminals as soon as they were notified about the attack~\cite{Lareau2020}.

\subsubsection{Magellan Healthcare Breach (Phoenix)}
During the Magellan Health (Phoenix) data breach, nearly $1,013,956$ individuals were affected. This attack was discovered in April $11^{th}$. The hacker cracked into the IT systems database through a phishing email sent to a company impersonating a client. This email was sent on April $5^{th}$. The hacker could access the whole database of the individual's sensitive personal information. The incident was immediately reported to the law enforcement authorities, including the FBI. The Fortune 500 company has given a statement saying that it has increased its security protocols after the breach~\cite{Davis}.
\subsubsection{Health Share of Oregon Breach (Portland)}
During the Health Share of Oregon attack, nearly $654,362$ individuals were affected. This attack happened on November $18^{th}$ when one of the laptops was stolen from the external vendor's office, i.e., Gridwork. This company provides non-emergency transportation for the Health Share of Oregon. The laptop was stolen on November $18^{th}$, but the healthcare organization was not informed until January. The laptop contained the information of the patient's name, address, phone numbers, date of birth, and SSN. However, the patient's health history was not there on the stolen laptop. As compensation, the company has given the clients(whose data has been compromised) free credit monitoring for a year~\cite{Terry}.

\subsubsection{Florida Orthopaedic Institute Data Breach (Tampa)}
During the Florida Orthopaedic breach, which occurred around April, $9^{th}$, nearly $640,000$ individuals were affected, making this the second-largest healthcare data breach in the year 2020~\cite{kumar2021decentralized}. The server was attacked, the organization encrypted the patient's data. After further investigation, it came to notice that all the data was stolen prior to the encryption. The data which was compromised contains all patients' sensitive data, including the social security number. Florida Orthopaedic Institute is unaware if any of this data was misused~\cite{Alder2020}. 
The reason for these attacks is still unidentified in some cases. According to the cyber security analysts, these attacks have been classified into various categories according to the threat level~\cite{kumar2021decentralized}.

\section{Different Types of Threat Levels of an attack}
The highest risks and the percentage by which these healthcare organizations are affected~\cite{ref1}:

\begin{itemize}
	\item Malicious network traffic: 72\% 
	\item Phishing: 56\%
	\item Vulnerable OS (high risk): 48\%
	\item Man-in-the-middle attack: 16\%
	\item Malware: 8\%
\end{itemize}

The medium risks and the percentage by which these healthcare organizations are affected~\cite{ref1}:
\begin{itemize}
	\item Configuration vulnerabilities: 60\%
	\item Risky hot spots: 56\%
	\item Vulnerable OS (all): 56\%
	\item Sideloaded applications: 24\%
	\item Unwanted or vulnerable application: 24\%
	\item Cryptojacking: 16\%
	\item Third-party app stores installed: 16\%
\end{itemize}

\subsection{WLAN attacks}
Wireless LAN (WLAN) implementations integrated with a wired network often are installed in hospital facilities, in large organizations down to the individual practice level, and in teaching and research hospitals~\cite{gibbs2008medical,haring2021review}. These systems are required to remain highly available as part of mission and life-critical systems and applications. They provide a high level of security, including interference notification and detection of rogue access points. Inspite of providing high level of security, they are not fully secure, there are two types of attacks. The WLAN attacks can be categorized as follows~\cite{hiltunen2008wlan}:
\begin{itemize}
	\item \textbf{Passive Attacks:} In the passive attacks, the hacker may analyze the WLAN packet traffic and capture the transmission methods. He gains an unauthorized user access to the network, but no modifications are made to the network
	\item \textbf{Active Attacks:} In the active WLAN attack, the hacker gains unauthorized access to the networks and he might modify the network by different methods like man-in-the middle attack, DoS, session hijacking.
	To prevent these attacks, they must follow the improved WLAN authentications and improved encryption methods which are discussed in the further sections.
\end{itemize}

\section{Solutions for the Data Breaches}
\subsection{Autonomic Computing}
Autonomic computing can automatically adapt to the different technologies to detect anomalies in the network and resolve the difficulties. Automatic Computing technology enables Distributed Systems to strategically learn novel attack patterns that help change and evolve the response evaluation algorithms, which will help detect zero-day attacks. Every autonomic computing system must contain eight key factors. 
\begin{enumerate}
	
	\item \textbf{Self Awareness:} The computing system should be aware of itself and in which state it is and its behavior. 
	\item \textbf{Self-configuration:} The computing system should be able to configure and reconfigure itself.
	\item \textbf{Self-optimization:} To enhance its operation, the computing system should be able to continuously optimized. 
	\item \textbf{Self-healing:} Whenever a failure occurs in the computing system, the system should be able to recover itself from the failures.
	\item \textbf{Self-Protection:} The system should be able to proactively detect any kind of cyberattacks and protect itself from those attacks.
	\item \textbf{Context awareness:} The computing system should be aware of its environment and act accordingly.
	\item \textbf{Open environment:} The system should function within its own environment and adapt to any kind of changes in the environment.
	\item \textbf{Estimated resource allocation:} The computing system should identify what kind of resources the system needs and optimize the resources.
\end{enumerate}

\subsection{Autonomic Security Management Framework (ASMF)}
\begin{figure}[h]
	\centering
	\includegraphics[width=9cm, height = 4.5cm]{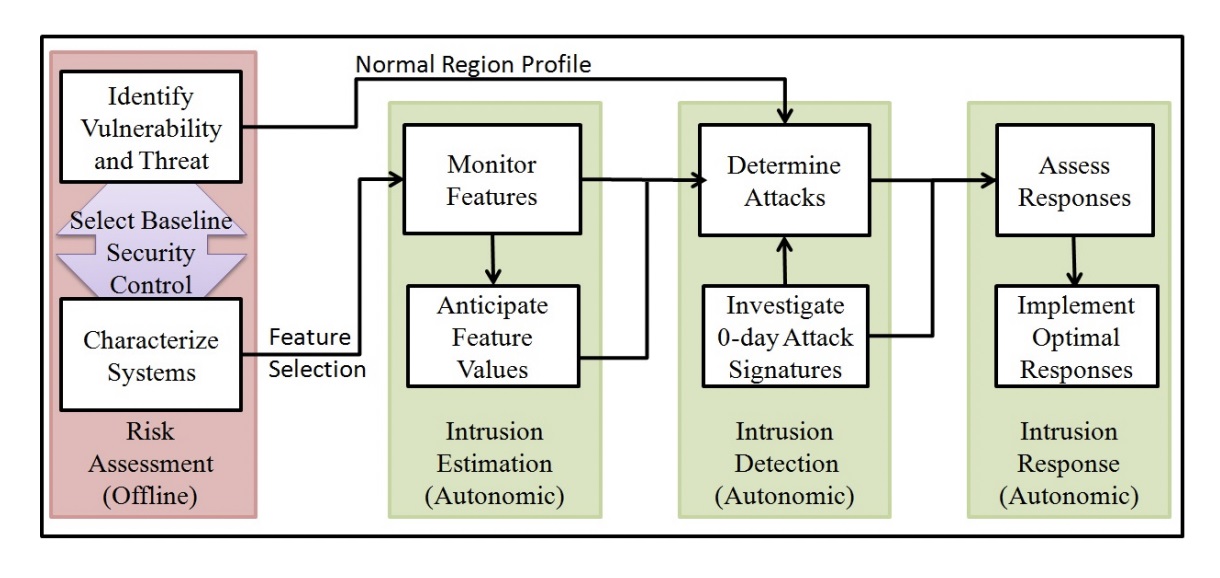}
	\caption{Autonomic Security Management Framework.\\
		Fig.~\ref{fig:fig 4} Alt Text: The security management framwork diargam tha consists of an offline risk assessment stage, followed by intrusion estimation, intrusion detection and intrusion response stages.}
	\label{fig:fig 4}
\end{figure}

This security framework contains Risk Assessment, Intrusion Detection, Intrusion Estimation, and Intrusion Response Module (REDR)~\cite{chen2016towards,cheng2008toward}. Figure~\ref{fig:fig 4} shows the autonomic security management framework diagram. This framework follows the guidelines of NIST and HITRUST. This can be an added advantage because of the less intervention of the humans. 

\textbf{Working of Each Module:}
\begin{enumerate}
	
	\item \textbf{Risk Assessment:} This is the first step in the REDR module, which helps assess its risk offline and helps the organization determine the risk of an attack and its	impact on the organization. This module is divided into two components which are Identifying threats and vulnerabilities and characterizing systems.
	
	\item \textbf{Intrusion Estimation:} System sensors monitor the selected features continuously. Upcoming attacks are anticipated by comparing the normal region and with the output of the system model. Early warnings are sent to the controller, and appropriate control mechanisms are executed to neutralize the attacks.
	
	\item \textbf{Intrusion Detection:} An intrusion detection system is a real-time monitoring event where the system detects any abnormality in the system's behavior. This anomaly detection technique which compares the real-time system performance, helps to detect unknown attacks.
	
	\item \textbf{Intrusion Response:} The IRS (Intrusion Response System) enables the self-protecting system to mitigate and regulate the system back to normal. This IRS is divided into two types, static type, and dynamic type. After the self-protecting systems perform the steps, sensors in feedback control theory are used to check if the selected features are continuously monitored. This will help to see if the system has returned to its normal running behavior. If any abnormal behavior is detected, the self-protection process mentioned will repeat itself.

\end{enumerate}
\section{Potential Solutions to Reduce Healthcare Data Breaches}

\subsubsection{Training and Educating the Employees}
One of the main reasons why many data breaches happen in healthcare is because of a lack of training and employee education in privacy and security. Cyber attacks can be stopped/reduced by educating the employees. In this way, health organizations can stay compliant with HIPAA and HITECH regulations. This training can create awareness about different phishing attacks and security risks to the employees. There are different training programs offered like SecurityIQ~\cite{howard2008test}, AwareEd~\cite{tjan1997leaders}, PhishSIM~\cite{dupuis2020clickthrough}. These are the platforms where it allows the employees to learn through hands-on learning, a phishing simulator.

\subsubsection{Implementation of Incident Response Plan}
There are six phases in IRP, they are:
\begin{itemize}
	\item \textbf{Preparation:} The IRP is going to prepare on how to handle a data breach in case it occurs.
	\item \textbf{Identification:} In this phase, the IRP is going to help in identifying the data breach and to identify if it is actually a threat or just a false alarm
	\item \textbf{Containment: } It disconnects the affected system with all the other networks and thus helping in minimizing the damage of the breach.
	\item \textbf{Eradication:} All the affected systems are removed and are replaced with the new ones.
	\item \textbf{Recovery:} After making sure that the affected systems are no longer a threat, they are put back into the network.
	\item \textbf{Lessons Learned:} This is the final phase where all the documentation regarding the incident is done, and analysis of the report is done to make sure that this type of breach does not occur again.
\end{itemize}

\section{Conclusion}
Data Breaches are on the rise in the health sector, and they keep increasing every year. Data breaches are highest in the health sector because of the high value of the Protected Health Information (PHI) on the dark web and the illegal black market. These data breaches happen for various reasons such as phishing, DoS attacks, and sometimes due to the system's human factor. These data breaches can affect the organization in many ways. Therefore, these cyberattacks can be minimized up to an extent by educating the employees and implementing the Incident Response Plan. This paper aims to study the reason for the recent data breaches and investigate the possible methodologies to improve the healthcare system's security and prevent it from future data breaches.

\bibliographystyle{ieeetr}

% Loading bibliography database
\bibliography{cas-refs}

\end{document}